\documentclass{iau} 
\usepackage{graphicx}

\title[Morphology of the Magellanic Clouds] 
{Morphology of stellar populations in the Magellanic Clouds using the VMC survey}

\author[Dalal El Youssoufi et al.]   
{Dalal El Youssoufi$^1$, Maria-Rosa L. Cioni$^1$, Cameron P. M. Bell$^1$, \\ Stefano Rubele$^{2,3}$, Florian Niederhofer $^1$, Gal Matijevic$^1$}
\affiliation{
	$^1$ Leibniz-Institut f\"ur Astrophysik Potsdam, An der Sternwarte 16, 14482 Potsdam, Germany \\[\affilskip]
	$^2$ Dip. di Fisica e Astronomia, Univ. di Padova, vicolo dell'Osservatorio 2, 35122 Padova, Italy \\[\affilskip]
	$^3$ INAF-Osservatorio Astronomico di Padova, vicolo dell'Osservatorio 5, 35122, Padova, Italy}

\pubyear{2018}
\volume{344}  
\jname{Dwarf Galaxies: From the Deep Universe to the Present}
\editors{S. Stierwalt \& K. McQuinn eds.}
\begin{document}
	\maketitle
	
	\begin{abstract}
		The Magellanic Clouds are nearby dwarf irregular galaxies that represent a unique laboratory for studying galaxy interactions.  Their morphology and dynamics have been heavily influenced by their mutual interactions as well as with their interaction(s) with the Milky Way. We use the VISTA near-infrared $YJK_{\mathrm{s}}$ survey of the Magellanic Clouds system (VMC) in combination with stellar partial models of the Large Magellanic Cloud (LMC), the Small Magellanic Cloud (SMC) and the Milky Way to investigate the spatial distribution of stellar populations of different ages across the Magellanic Clouds. In this contribution, we present the results of these studies that allow us to trace substructures possibly related to the interaction history of the Magellanic Clouds.
		
	\end{abstract}

	\section{Introduction}
	Located at approximately 50 and 60 kpc, the Magellanic Clouds (MCs) represent the nearest interacting pair of galaxies to the Milky Way. They have been targets of intensive research for many years, making them an unparalleled  benchmark for studying galaxy interactions in a near-field cosmological context. Our view of the MCs has changed rapidly during the last decade: proper motion measurements suggest that the MCs may be on their first infall to the Milky Way (e.g. \cite[Kallivayalil et al. 2006]{Kallivayalil2006}), new satellites of the MCs were discovered (e.g. \cite[Torrealba et al. 2016]{Torrealba2016}) and new substructures were found in their outskirts (e.g. \cite[Mackey et al. 2018]{Mackey2018}). 
	
	The LMC is an almost face on, gas-rich galaxy with an offset bar. It is known for its non-planar structure (e.g. \cite[Nikolaev et al. 2004]{Nikolaev2004}) characterized by warps and twists (e.g. \cite[Olsen \& Salyk 2002]{Olsen2002}; \cite[Choi et al. submitted]{Choi}) and for its distinct multiple spiral arms (e.g. \cite[deVaucouleurs \& Freeman. 1975]{deVaucouleurs1975}; \cite[Besla et al. 2016]{Besla2016}). The structure of the SMC appears ellipsoidal and strongly elongated along the line-of-sight (e.g.  \cite[Scowcroft et al. 2016]{Scowcroft2016};  \cite[Jacyszyn-Dobrzeniecka et al. 2016]{Jacyszyn2016}). The SMC is known for its less pronounced bar and eastern extension, also named the Wing. 
	Young and old stellar populations display different spatial distributions across the MCs. In the LMC, young stars exhibit a rather irregular structure characterized by spiral arms and tidal features, while older stars tend to be more smoothly and regularly distributed (e.g. \cite[Cioni et al. 2000]{Cioni2000}; \cite[Nikolaev \& Weinberg 2000]{Nikolaev2000}). Young populations in the SMC are concentrated in the central parts and in the Wing, while the older populations can be designated as a spheroid or ellipsoid (e.g. \cite[Cioni et al. 2000]{Cioni2000}; \cite[Zaristky et al. 2000]{Zaritsky2000}). The morphology of the MCs can be considered as a fossil record of their interaction history and its study deems itself important as tidal forces caused structural changes in the galaxies. 
	
	\section{Stellar population selection and data analysis}
	The VISTA near-infrared survey of the Magellanic Clouds system (VMC, \cite[Cioni et al. 2011]{Cioni2011}) offers the deepest $YJK_{\mathrm{s}}$ band photometry across the MCs to date, reaching a depth of  $K_{\mathrm{s}}$=20.3 (Vega) at S/N=10 with a sensitivity corresponding to the bottom of the red giant branch population.
	Our aim is to derive the morphology of the MCs using different stellar populations. We have used the ($J-K_{\mathrm{s}}, K_{\mathrm{s}}$) colour-magnitude diagram (CMD) in combination with stellar partial models of the LMC, SMC, and of the Milky Way to select different classes of objects. We used CMD regions defined in \cite[Cioni et al. (2014)]{Cioni2014} for the LMC while for the SMC, we shifted these regions both in colour and magnitudes to take into account differences in mean distance and metallicity of the stellar populations, as in \cite[Cioni et al. (2016)]{Cioni2016}. The exact location of the CMD regions is based on the analysis of the star formation history by \cite[Rubele et al. (2012)]{Rubele2012}.  Furthermore, in order to assess the completeness of our aperture photometry, we used artificial star tests performed on point spread function photometry catalogues and found that it shows a good performance even in the crowded parts of the galaxies. In addition, we also used stellar partial models to determine the Milky Way contamination in our regions and found that regions F and H have the highest contamination levels (over 70\%) followed by regions G and I ($\sim$ 15\%). We did not correct our data for reddening as the dust content of the Magellanic Clouds is low, $E(V-I)\sim0.06$ mag (\cite[Haschke et al. 2011]{Haschke2011}) on average corresponding to an absorption A$_J=0.04$ and A$_{Ks}=0.02$ mag using the Cardelli reddening law. 
	
	\begin{figure}[h!]
		\begin{center}
			\includegraphics[scale=0.37]{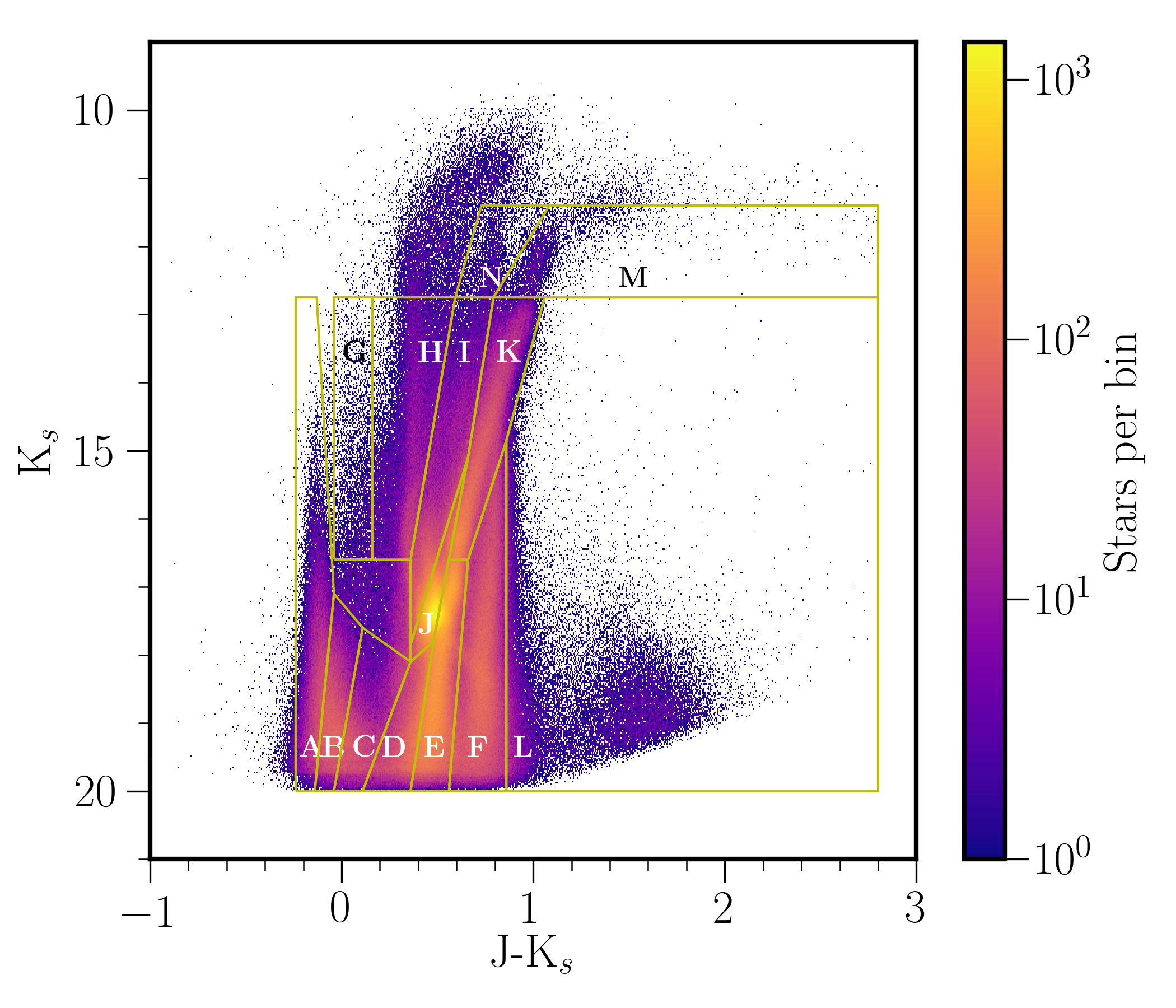} 
			\includegraphics[scale=0.37]{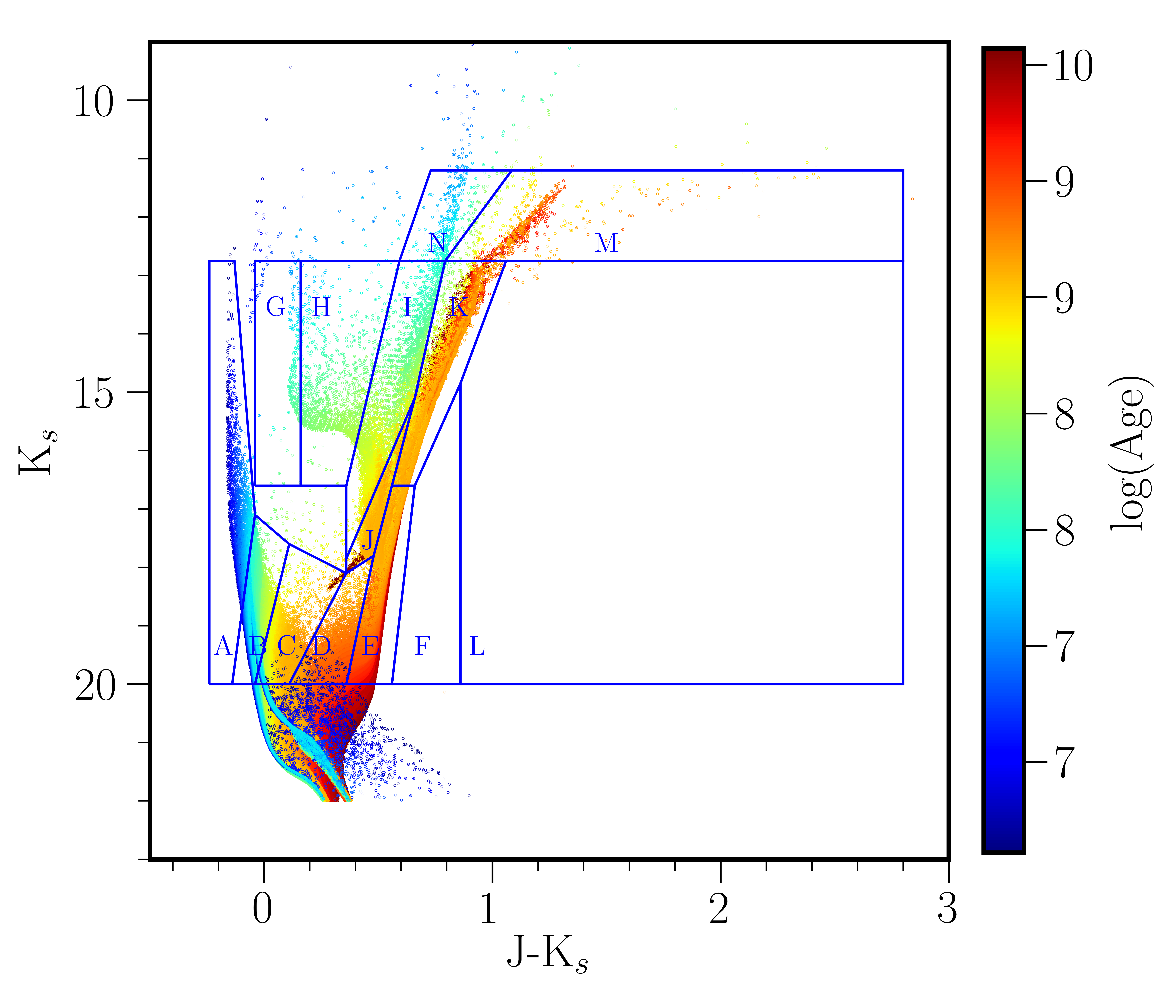}		
			\caption{\textbf{Left:} Near-infrared ($J-K_{\mathrm{s}}, K_{\mathrm{s}}$) Hess diagram of the SMC. The colour scale indicates the stellar density in logarithmic scale while the yellow boxes indicate the boundaries of our regions; the bin size in $K_{\mathrm{s}}$ is 0.0055 mag and in $J-K_{\mathrm{s}}$ it is 0.0220 mag. \textbf{Right:} Simulated ($J-K_{\mathrm{s}}, K_{\mathrm{s}}$) CMD illustrating stellar populations of the SMC corresponding to a wide range of ages as shown in the figure and metallicites ([M/H]=-1.68 dex to [M/H]=-0.55 dex).}
			\label{fig1}
		\end{center}
	\end{figure}
	
	\section{Morphology of the Magellanic Clouds}
\begin{figure}[h!]
		\begin{center}
			\includegraphics[scale=0.24]{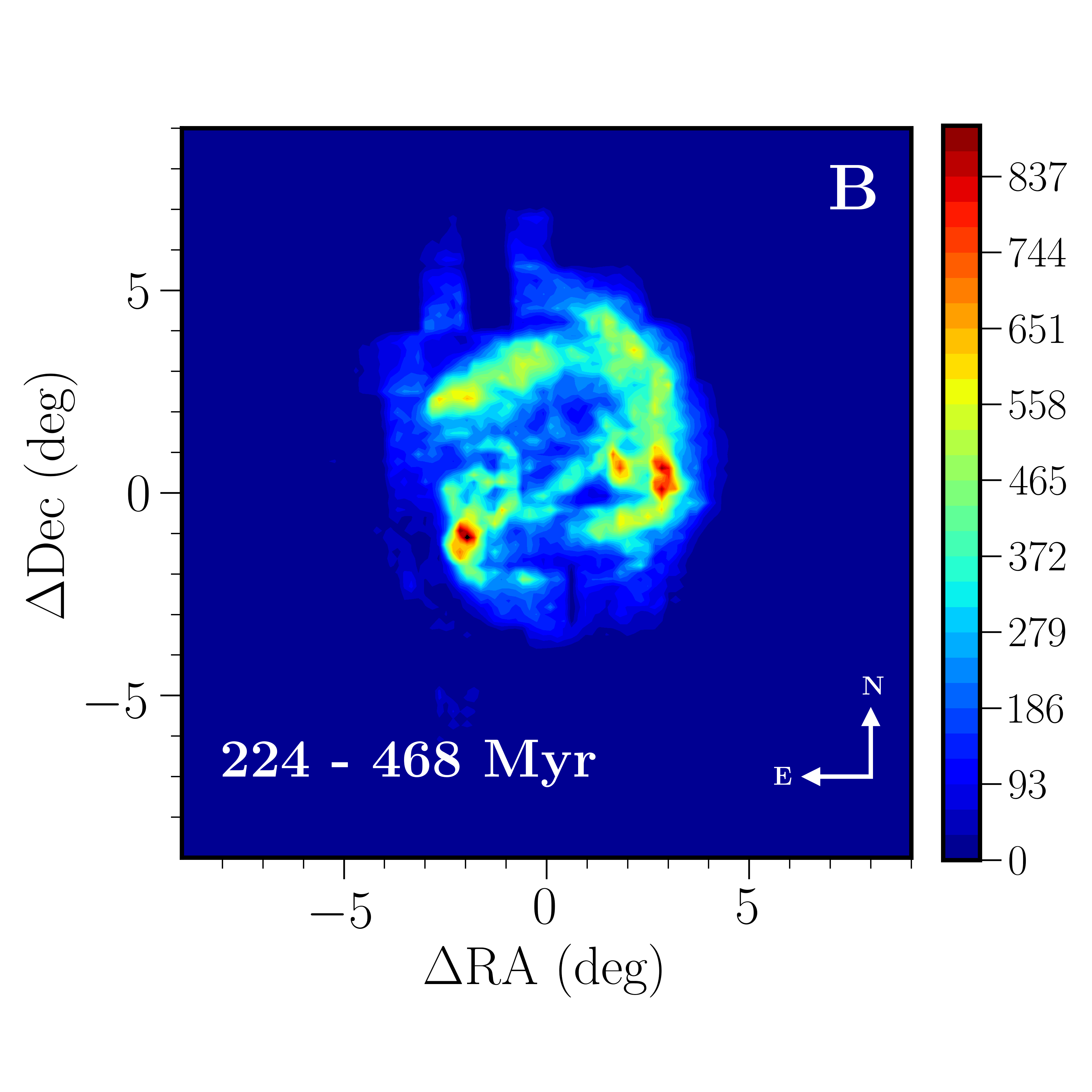} 
			\includegraphics[scale=0.24]{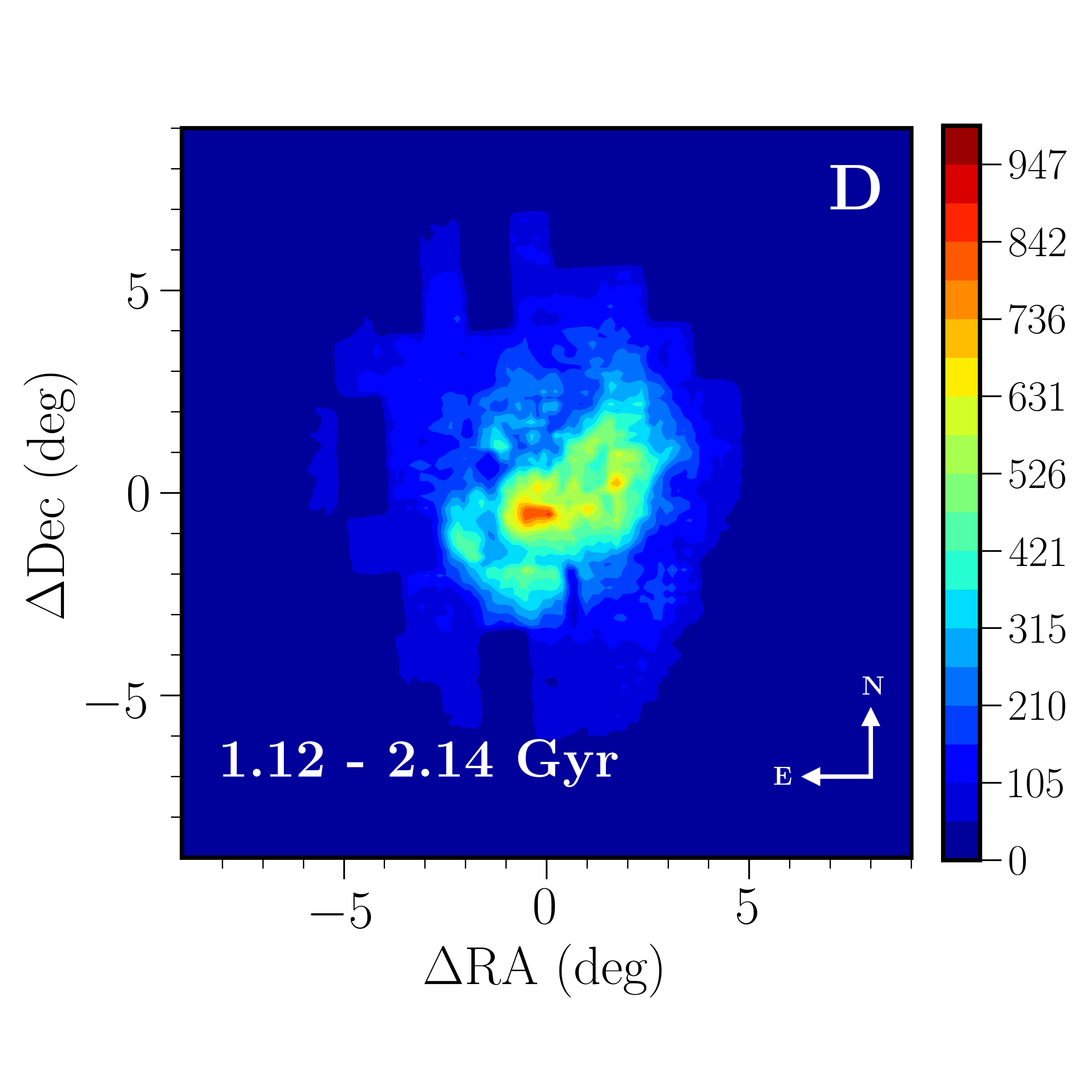}
			\includegraphics[scale=0.24]{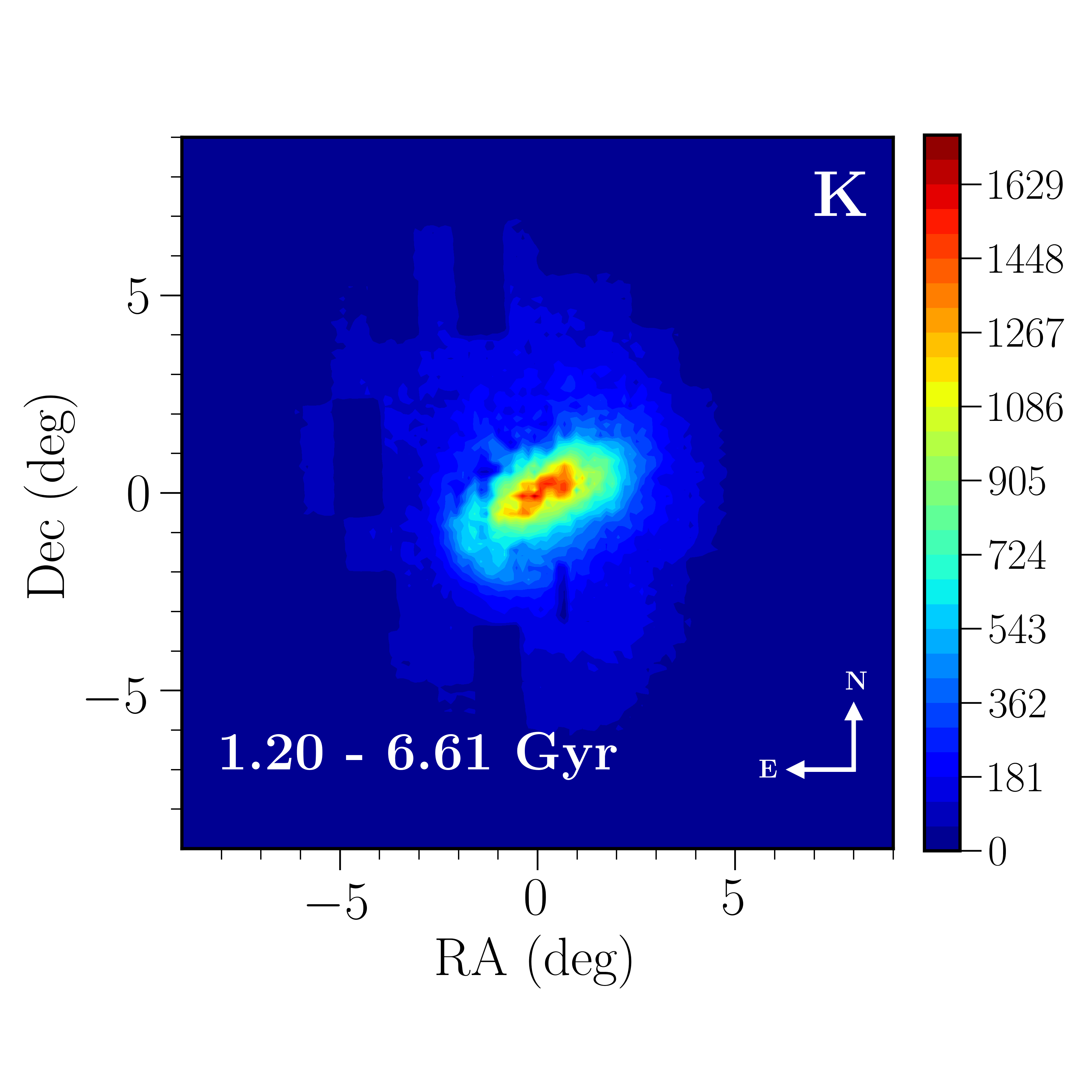}\\
			\includegraphics[scale=0.24]{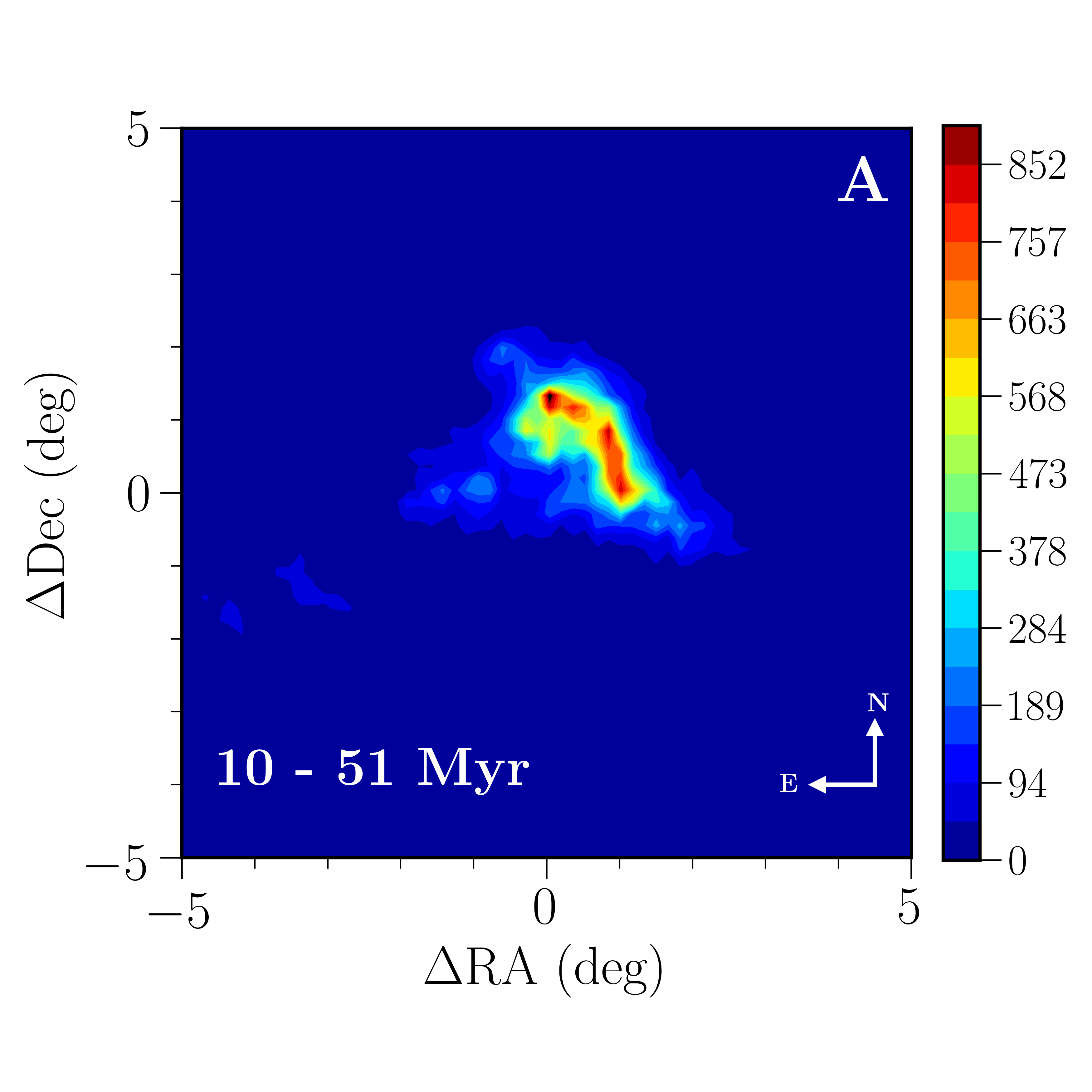}
			\includegraphics[scale=0.24]{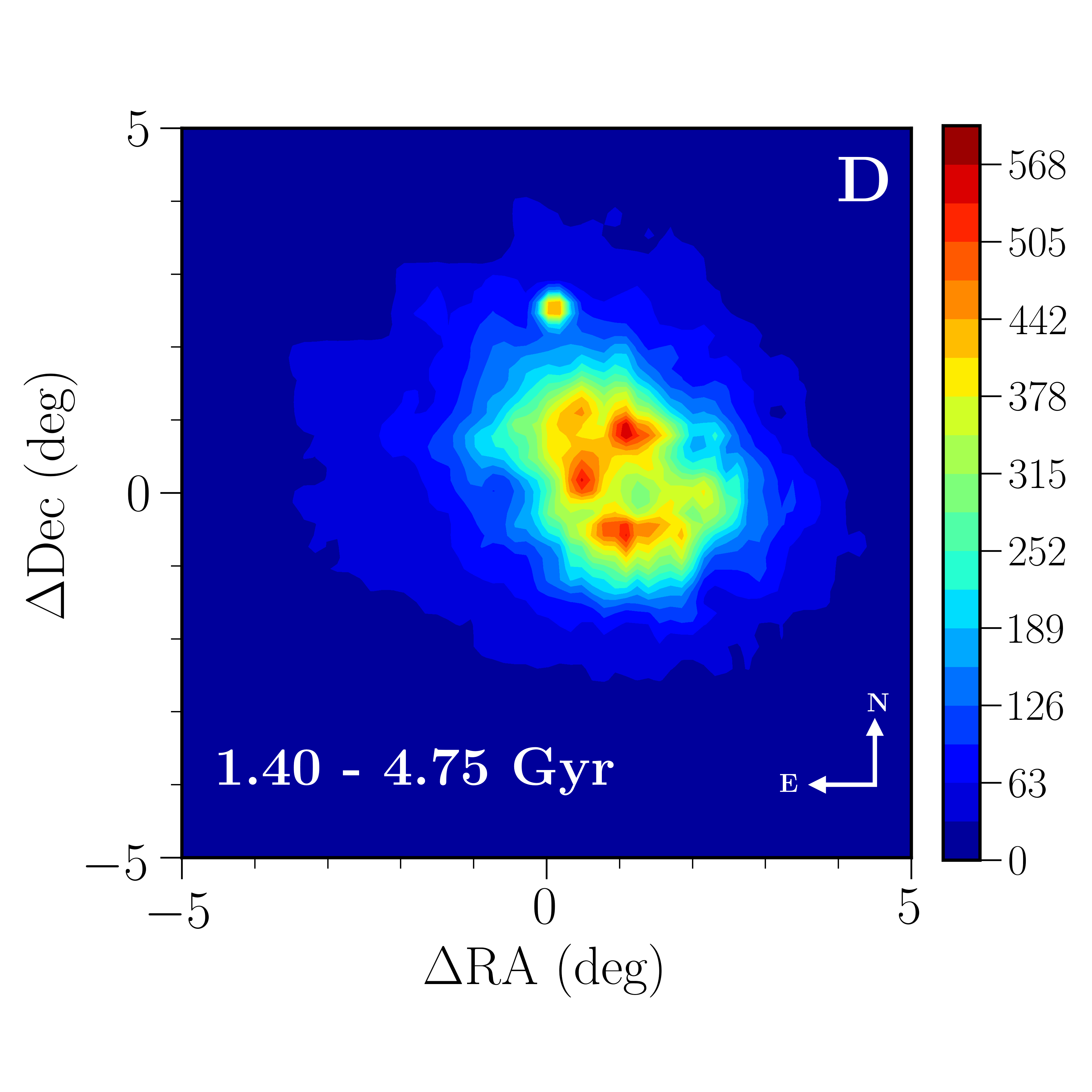}
			\includegraphics[scale=0.24]{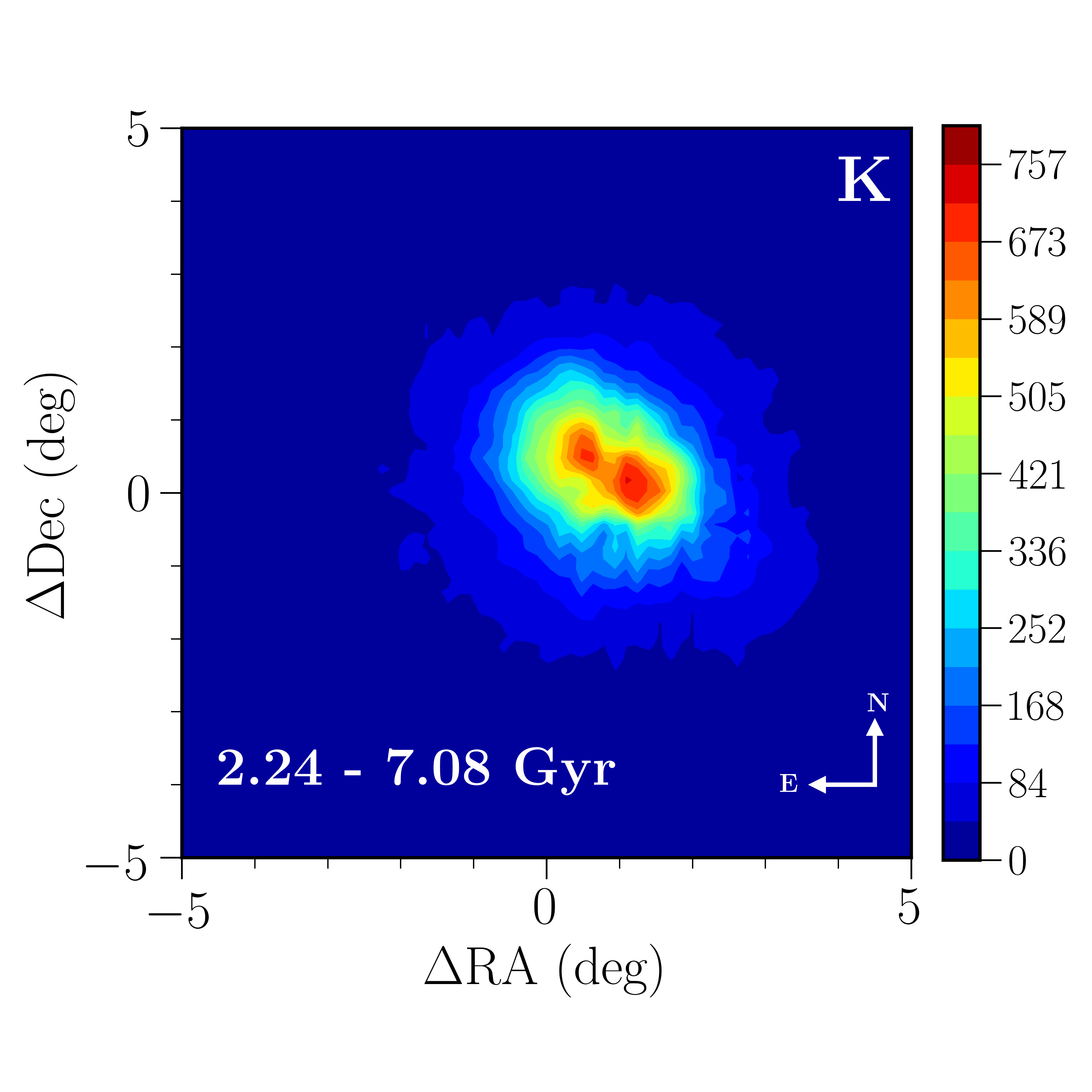} 
			\caption{Examples of stellar density/contour maps of the LMC \textbf{(top)} and SMC \textbf{(bottom)}, for young \textbf{(left)}, intermediate-age \textbf{(middle)} and  old \textbf{(right)} stellar populations. The colour bar shows the contour levels outlining the number of stars per bin, the bin size is 0.03 deg$^2$. The projection origin for the LMC and SMC are respectively set to be at (RA$_{0}~$,~Dec$_{0}$) = (81.00$^{\circ}$ , -69.73$^{\circ}$) and (RA$_{0}~$,~Dec$_{0}$) = (13.05$^{\circ}$ , -72.82$^{\circ}$).}
			\label{fig2}
		\end{center}
	\end{figure}

Traced by different stellar populations, the morphology of the MCs shows different properties, the detailed structures in the central regions of our maps are characterised for the first time at the spatial resolution of 0.13 kpc and 0.16 kpc for the LMC and SMC, respectively. Figure \ref{fig2} provides examples of morphological maps of young main sequence stars (region A and/or B), subgiant and/or main sequence stars (region D) and red giant branch stars (region K) across the MCs.

Young main sequence stars in the LMC exhibit coherent structures. Two star forming regions are outlined as over densities, Shapely's constellation III ($\Delta$RA=-1 deg, $\Delta$Dec=2.5 deg) and 30 Doradus ($\Delta$RA=0.5 deg, $\Delta$Dec=0.5 deg), these features seem to grow dimmer as we progress in age. The bar is traced by a thin structure with no central nucleus, three major overdensities are visible, one at either the East and West end of the classical bar ($\Delta$RA=-2 deg to $\Delta$RA=2 deg), in addition to a third western density ($\Delta$RA=2 deg), a clear break between the classical bar and the western density is also visible, this break becomes more inconspicuous as the population becomes younger. Main sequence stars also trace the distinct multi-arm structure of the LMC, we found a clear connection between the southern tip of the South-West spiral arm ($\Delta$RA=1 deg, $\Delta$Dec=-0.5 deg) and the bar becomes more enhanced with age. Other young populations such as supergiants and asymptotic giant branch stars trace similar substructures. The overall structure of the galaxy, as traced by intermediate-age subgiant and/or main sequence stars, appears to be more regular with no significant spiral arms except for the South-West arm. The population has one major overdensity while the bar itself is inconspicuous. The overall morphology of these stars resembles that traced by red clump stars, but with a less dense nucleus. The spatial distribution of the red giant branch population is smooth and regular while the bar is more prominent with an overdensity at its centre resembling that of other old tracers such as from RR Lyrae stars.

The SMC's young main sequence stars show an asymmetric distribution with an irregular bar and an eastern extension. The presence of the Wing is most evident is these populations ($\Delta$RA=-1 deg, $\Delta$Dec=0 deg), protuberances possibly related to tidal interaction events are visible in the North ($\Delta$RA=1 deg, $\Delta$Dec=2 deg) and the South ($\Delta$RA=-2 deg to $\Delta$RA=3 deg, $\Delta$Dec=-0.3 deg) of the bar. The intermediate-age population showcases the transition from an asymmetric irregular structure to a spheroidal one, several clumps are visible outlining the partially smooth distribution (e.g. $\Delta$RA=0.5 deg, $\Delta$Dec=-0.3 deg). On the other hand, the old stellar population of the SMC shows a smooth and regular distribution which is typical for a spheroidal body.
	
	\section{Conclusions}
In this contribution, we utilised near-infrared photometric data from the VMC survey in combination with stellar partial models of the MCs and Milky Way to carry out a detailed and comprehensive study of the morphological properties of the MCs. We produce several stellar density/contour maps at the spatial resolution of 0.13 kpc and 0.16 kpc for the LMC and SMC, respectively, for various stellar populations, we find that the multi-arm structure of the LMC grows more enhanced with age in main sequence stars, most of the young populations show no clear separation between the classical bar and the western density of the LMC except for the oldest main sequence stars. Our morphological maps of the SMC highlight the transition of the SMC morphology as a function of time from a spheroidal system to an asymmetric and an irregular one.
	
\begin{acknowledgements}
		We thank the Cambridge Astronomy Survey Unit (CASU) and the Wide Field Astronomy Unit (WFAU) in Edinburgh for providing calibrated data products under the support of the Science and Technology Facility Council (STFC). This project has received funding from the European Research Council (ERC) under European Union's Horizon 2020 research and innovation programme (grant agreement No 682115). This study is based on observations obtained with VISTA at the Paranal Observatory under program ID 179.B-2003.
\end{acknowledgements}


\begin{thebibliography}{}
		
		\bibitem[Besla \etal\ (2007)]{Besla2007}
		{Besla, G., Kallivayalil, N., Hernquist, L., et al.} 2007, 
		\textit{ApJ}, 668, 949
		
		\bibitem[Besla \etal\ (2016)]{Besla2016}
		{Besla, G., Martinez-Delgado, D., van der Marel, R. P., et al.} 2016, 
		\textit{ApJ}, 825, 20
		
		\bibitem[Choi \etal\ (submitted]{Choi}
		{Choi, Y., Nidever, D. L., Olsen, K., et al.} submitted, arXiv: 1805.00481 
		
		\bibitem[Cioni \etal\ (2000)]{Cioni2000}
		{Cioni, M.-R. L., Habing, H. J., \& Israel, F. P.} 2000, 
		\textit{A\&A},  358, L9
		
		\bibitem[Cioni \etal\ (2011)]{Cioni2011}
		{Cioni, M.-R. L., Clementini, G., Girardi, L., et al.} 2011, 
		\textit{A\&A}, 527, A116
		
		\bibitem[Cioni \etal\ (2014)]{Cioni2014}
		{Cioni, M.-R. L., Girardi, L., Moretti, M. I., et al.} 2014, 
		\textit{A\&A}, 562, A32
		
		\bibitem[Cioni \etal\ (2016)]{Cioni2016}
		{Cioni, M.-R. L., Bekki, K., Girardi, L., et al.} 2016, 
		\textit{A\&A}, 586, A77
		
		\bibitem[de Vaucouleurs \& Freeman\ (1972)]{deVaucouleurs1975}
		{de Vaucouleurs, G. \& Freeman, K. C} 1975, 
		\textit{Vistas in Astron.}, 14, 163
		
		\bibitem[Haschke \etal\ (2011)]{Haschke2011}
		{Haschke R., Grebel E. K., Duffau S..} 2011, 
		\textit{AJ}, 141, 158
		
		\bibitem[Jacyszyn-Dobrzeniecka \etal\ (2016)]{Jacyszyn2016}
		{Jacyszyn-Dobrzeniecka, A. M., Skowron, D. M., Mroz, P., et al.} 2016, 
		\textit{AcA}, 66, 149
		
		\bibitem[Kallivayalil \etal\ (2006)]{Kallivayalil2006}
		{Kallivayalil, N., van der Marel, R. P., Alcock, C., et al.} 2006, 
		\textit{ApJ}, 668, 949
		
		\bibitem[Mackey \etal\ (2018)]{Mackey2018}
		{Mackey D., Koposov S., Da Costa G., Belokurov V., Erkal D., Kuzma P.,} 2018, 
		\textit{ApJ}, 858, L21
		
		\bibitem[Nikolaev \etal\ (2000)]{Nikolaev2000}
		{Nikolaev, S. \& Weinberg, M. D.} 2000, 
		\textit{ApJ}, 542, 804
		
		\bibitem[Nikolaev \etal\ (2004)]{Nikolaev2004}
		{Nikolaev, S., Drake, A. J., Keller, S. C., et al.} 2004, 
		\textit{ApJ}, 601, 260
		
		\bibitem[Olsen \& Salyk\ (2002)]{Olsen2002}
		{Olsen, K. A. G. \& Salyk, C.} 2002, 
		\textit{AJ}, 124, 2045
		
		\bibitem[Rubele \etal\ (2012)]{Rubele2012}
		{Rubele, S., Kerber, L., Girardi, L., et al.} 2012, 
		\textit{A\&A}, 537, A106
		
		\bibitem[Scowcroft \etal\ (2016)]{Scowcroft2016}
		{Scowcroft, V., Freedman, W. L., Madore, B. F., et al.} 2016, 
		\textit{ApJ}, 816, 49
		
		
		\bibitem[Torrealba \etal\ (2016)]{Torrealba2016}
		{Torrealba, G., Koposov, S. E., Belokurov, V., \& Irwin, M.} 2000, 
		\textit{MNRAS}, 459, 2370
		
		
		\bibitem[Zaritsky \etal\ (2000)]{Zaritsky2000}
		{Zaritsky, D., Harris, J., Grebel, E. K., \& Thompson, I. B.} 2000, 
		\textit{ApJ}, 534, L53
		
		
		
	\end{thebibliography}
\end{document}